\documentclass[pre,twocolumn]{revtex4}

\usepackage{graphicx}
\usepackage{dcolumn}
\usepackage{bm}
\usepackage[dvips]{color}

\def\mrm{\mathrm}
\def\mbf{\mathbf}

\def\etal{{\it et al. }}

\def\qq{\mbf{q}}

\def\phis{\phi_\mathrm{solid}}
\def\phif{\phi_\mathrm{fluid}}
\def\phihc{\phi_\mathrm{hc}}

\def\mrm{\mathrm}
\def\mbf{\mathbf}

\def\qq{\mbf{q}}

\def\s0{\langle s_0 \rangle}

\def\rbar{\overline{r_i}}

\begin{document}


\preprint{APS/123-QED}

\title{
Dynamical Study of Polydisperse Hard-Sphere System
}

\author{Tomoaki Nogawa}
\email{nogawa@serow.t.u-tokyo.ac.jp}
\author{Nobuyasu Ito}
\affiliation{%
Department of Applied Physics, 
The University of Tokyo, 7-3-1 Hongo, Bunkyo-ku, Tokyo 113-8656, Japan
}%
\author{Hiroshi Watanabe}
\affiliation{%
Supercomputing Division, Information Technology Center, 
University of Tokyo, 2-11-16 Yayoi, Bunkyo-ku, Tokyo 113-8658, Japan
}%

\begin{abstract}

We study the interplay between 
the fluid-crystal transition and the glass transition 
of elastic sphere system with polydispersity 
using nonequilibrium molecular dynamics simulations. 
It is found that 
the end point of the crystal-fluid transition line, 
which corresponds to the critical polydispersity 
above which the crystal state is unstable, 
is on the glass transition line. 
This means that crystal and fluid states at the melting point 
becomes less distinguishable as polydispersity increases 
and finally they become identical state, i.e., 
marginal glass state, at critical polydispersity. 

\end{abstract}

\maketitle

\begin{figure}[t]
\begin{center}
\includegraphics[trim=20 260 0 -225,scale=0.310,clip]{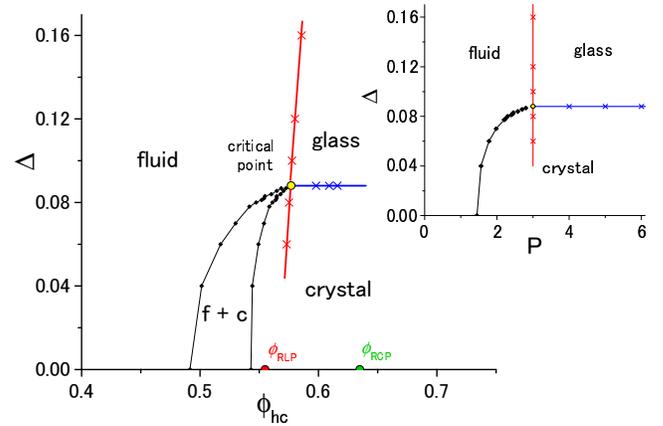}
\end{center}
\vspace{-5mm}
\caption{\label{fig:PD}
(color online) 
Phase diagram showing the polydispersity vs packing fraction 
(or pressure, inset) plane. 
There are three equilibrium phases: fluid, crystal, 
fluid-crystal coexistence. 
The boundary between the fluid and the disordered solid 
indicates a dynamical transition. 
}
\end{figure}

\section{Introduction}

Recently, jammed (amorphous) solid has attracted more attention 
in two aspects; 
1) the origin of rigidity which fluid lacks, and 2) whether some 
essential differences exists between the jammed solid and crystal 
except for the positional order.
The former corresponds to the long standing problem 
of glass transition \cite{Debenedetti01} 
and jamming transition \cite{OHern03}, 
and the latter is related to the dense packing problem \cite{Torquato00}.
Since whether the solid has or lacks the positional order
is complexly related to properties of materials,
it is difficult to find which factor is important.
Systematic study is, however, possible in polydisperse particle system, 
which consists of multiple ingredients with various sizes or shapes. 
It is empirically known that binary fluid mixture, 
with more than 10\% size-dispersion, often exhibits glass transition
\cite{Auer01} while monodisperse simple fluid 
exhibits crystallization transition.
Therefore, polydispersity is one of the most important factors
for the glass transition.
While the system may involve the glass transition 
for high enough polydispersity
and the simple crystallization for low enough polydispersity,
there remains a gap in knowledge between these two regimes.
In order to study the effect of polydispersity, accurate control of 
polydispersity is required, which is difficult in experiments.

Here, we consider the polydisperse hard-sphere (HS) system, 
which is one of the simplest models to exhibit 
both of fluid and crystal phases. 
While it is well known that monodisperse HS system takes 
a first order melting/crystallization transition, 
so-called the Alder transition, 
by increasing/decreasing pressure or density \cite{Wood57, Alder57}, 
much attention recently has been attracted to the problem;
how polydispersity affects this transition. 
Most remarkable finding is that 
the discontinuity at the melting point, 
i.e., density gap between the fluid and the crystal, 
decreases as the strength of polydispersity increases 
and finally vanishes at a certain critical point 
\cite{Ito96, Lahijany97, Vermohlen95} 
(see Fig.~\ref{fig:PD}). 
This is similar to the well-known liquid-gas criticality 
in systems with attractive interactions 
but there are some differences. 
The fluid and crystal states are distinguished 
by their spatial periodicity and fluidity, 
in addition to their density. 
Therefore, there is another transition(s) 
corresponding to the two properties in the supercritical region. 
When periodic order is not established even after fluidity is lost, 
there must be an intermediate phase, 
which is considered the glass phase \cite{Chaudhuri05}. 
The transition from fluid to glass is considered 
to be dynamical transition 
corresponding to ergodicity breaking\cite{vanMegen93, Zaccarelli09}.

The phase behavior of a polydisperse HS system 
still remains under discussion. 
Bartlett and Warren studied polydisperse systems 
using a density functional theory (DFT) 
and claimed that the thermodynamic function 
does not have a singularity at the point of equal concentration 
and that the first-order transition line is extended to high-density region 
to surround the crystal phase \cite{Bartlett99}. 
Furthermore, Fasolo \etal pointed out the importance of fractionation; 
segregation into multiple crystals. 
Each crystal has different mean radius 
and relative dispersion is small inside it. 
By considering the free energies of mixed states, 
a lot of coexisting phases appears 
and the phase diagram becomes very complicated\cite{Fasolo03}. 
Although such fractionated states is reasonable in thermal equilibrium state, 
these phases are not observed in experiments or numerical simulations. 
One reason is that the system is glassy in the regime 
where fractionation is predicted. 
Therefore diffusion of particles over long distance, 
which is necessary for segregation, is suppressed.

In the present paper, 
we don't treat fractionation 
but consider long surviving homogeneous state 
including both of equilibrium and nonequilibrium ones. 
Especially, we discuss the relation 
between the fluid-crystal transition for small dispersity  
and the fluid-glass transition for large dispersity.
We perform nonequilibrium molecular-dynamics (MD) simulations, 
which is not only useful to study nonequilibrium dynamics 
but also gives clue to reveal equilibrium property. 
We study the fluid-crystal transition in equilibrium 
for low dispersity by nonequilibrium simulation. 
On this aspect, 
a number of numerical studies on polydisperse hardcore systems 
has been previously reported. 
But these have been highly restricted to two-dimensional hard-disk systems 
\cite{Strandburg88, Watanabe05}. 
Since two-dimensional systems show peculiar properties 
owing to the low dimensionality, 
the study of three-dimensional systems is necessary. 
On the other hand, it is difficult to perform simulations 
with sufficiently large linear dimensions in three dimensions, 
thus finite size effect often makes the conclusions ambiguous. 
Nonequilibrium analysis without time-consuming equilibration 
makes large-scale simulations possible. 


Let us denote the contents of the present paper.
In the next section, 
the detail of the model and method of a numerical simulation is explained. 
From section III to V, 
we investigate three transitions among 
fluid, crystal and glass states 
to obtain the nonequilibrium phase diagram 
shown in Fig.~\ref{fig:PD}.
The final section is devoted for the concluding remarks.

\section{Model: Hard elastic spheres}

We perform MD simulations of elastic spheres 
with a fixed number of particle $N$, 
temperature $T$ and pressure $P$ 
using the Nos\'e-Hoover method \cite{Nose84,Hoover85} 
and the Parinello-Rahman method \cite{Parrinello81}. 
The reason we did not employ 
the standard event-driven simulation of HSs \cite{Alder59} 
is that pressure control, 
which is essential in the following analysis of first order transition, 
cannot be implemented efficiently to this method.
Since hard-sphere system is widely accepted 
as one of reference models for solid-fluid transitions, 
we estimate hard-limit of elastic modulus by extrapolation, 
which is described later. 
Hereafter, we use the units with which 
temperature $k_B T = 1$, mean radius $\rbar=1$, 
and the mass of the particle to be $m_i=1$. 

Polydispersity is introduced by a uniform distribution of particle radii.
The strength of polydispersity 
is measured by the standard deviation, 
$\Delta = \sqrt{\overline{ (r_i - 1 )^2} }$, 
where $r_i$ is the radius of particle $i$ and 
$\overline{\cdots}$ denotes the average over all particles. 
It is known that the quantitative properties of a polydisperse system 
are well described only by $\Delta$ 
and that the detailed form of the distribution function of $r_i$ 
is irrelevant when polydispersity is not too strong \cite{Ito96}.

The interaction between contacting particles, $i$ and $j$, 
is given by the Hertzian contact potential, 
$E_0 [ |\qq_i - \qq_j| - (r_i+r_j) ]^{5/2}$, 
where $\qq_i$ is the position of particle $i$. 
The interaction energy equals zero 
when $|\qq_i - \qq_j| > r_i+r_j$. 
The system becomes a true HS 
when Young's modulus $E_0$ approaches infinity. 
Young's modulus is set to $E_0=10^4 - 10^7$. 
Since we use finite values of Young's modulus,
the particles are allowed to overlap 
to make the effective radius and density smaller.
In order to correct this effect, we consider the effective hardcore
packing fraction $\phi_\mrm{hc}$, which corresponds to the density
of the system with an infinite Young's modulus.
By considering the equipartition of energy, 
the overlap length of particles is proportional to $E_0^{-2/5}$.
The packing fraction of the corresponding hardcore system $\phi_\mrm{hc}$
is therefore expected to be
$\phi_\mrm{hc} = (4\pi/3V) \sum_i 
\left( r_i - c_0 E_0^{-2/5} \right)^3$
with a calibration constant $c_0$.
This constant is determined to be $c_0 = 0.48$ 
by performing preliminary simulations, 
with which we confirmed that the extrapolated state equation 
exhibits good agreement with the result 
of event-driven simulation of the true HS system. 
This correction is used throughout the letter. 
For example, $\phi_\mrm{hc}$ is 0.6\% smaller than 
$(4\pi/3V) \sum_i r_i^3$ for $E_0=10^6$.

We adopt two types of initial particle configuration; 
FCC and random configurations. 
The radii of particles are assigned randomly 
in accordance with the distribution mentioned above 
and independently of the positions. 
For random initial state, 
we perform simulations with overdamped dynamics 
before integrating Hamilton's equations of motion 
until the maximum kinetic energy of the particles decreases below $200$ 
to avoid the rapid acceleration of strongly coalesced particle pairs. 
After that, the initial velocities are randomly assigned 
by the Boltzmann distribution. 

The observed quantities discussed below are the data 
for $E_0=10^6$ averaged over 4 samples with $N=55296$, 
unless otherwise stated. 
We confirm that our conclusions do not change 
in a larger system with $N=131072$. 
Time integration is performed 
by the fourth-order predictor-corrector method 
using a discrete time step of $\Delta t=0.0004-0.02$ 
and typical number of integration step is $4 \times 10^6$.

\section{Crystal melting transition}

First, we analyze the polydispersity dependence 
of the fluid-crystal transition 
and clarify the existence of the predicted critical point 
$(P_c, \Delta_c)$. 
The order parameter corresponding to this criticality 
is the density gap $\delta \phi$ 
between the bistable phases at the melting point. 
This is calculated by a two-step simulation. 
As a first step, we determine the melting pressure $P_m(\Delta)$ 
for a given $\Delta(<\Delta_c)$ from the nonequilibrium analysis 
discussed later. 
After that, we observe the packing fractions 
of the bistable states, 
$\phif \left( \Delta, P_m(\Delta) \right)$ and 
$\phis \left( \Delta, P_m(\Delta) \right)$, 
individually by performing equilibrium simulations 
with both fluid (random) and crystal(FCC) initial conditions. 
The packing fraction of the fluid/crystal at the melting point 
gives the lower/upper bound of the coexisting phase 
at a fixed $\phi$ condition (see Fig.~\ref{fig:PD}) 
and its width is 
$\delta \phi(\Delta) = 
\phi_\mathrm{solid}\left( \Delta, P_m(\Delta) \right) - 
\phi_\mathrm{fluid}\left( \Delta, P_m(\Delta) \right)$. 

\begin{figure}
\begin{center}
\includegraphics[trim=40 400 70 -225,scale=0.350,clip]{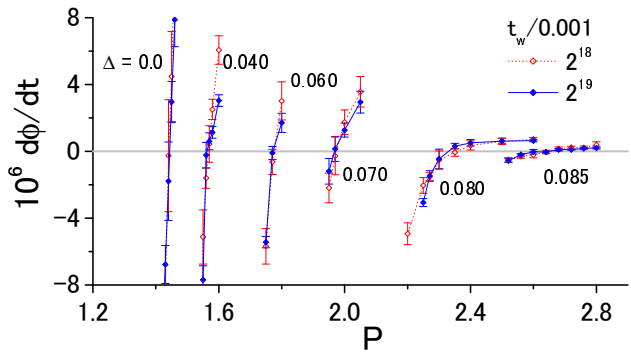}
\end{center}
\vspace{-5mm}
\caption{\label{fig:dphidt-P}
(color online) 
Time derivatives of the mean density is plotted with respect to pressure 
for $\Delta = 0.0 - 0.085$. 
The derivative $d \phi/ dt$ is approximated by 
$\left[ \phi(2t_w) - \phi(t_w) \right]/t_w$ 
with $t_w/0.001 = 2^{19}$ and $2^{20}$.
}
\end{figure}

To determine the melting point, 
we observe the nonequilibrium relaxation 
from the mixed initial state \cite{Ozeki07, Ozeki03a}; 
a half of the cubic space is occupied by the crystal 
and the remaining part is occupied by the random packing (fluid). 
Thus the two regions are separated by a flat interface, 
which is perpendicular to the (100)-direction of the FCC structure 
at time $t=0$. 
As $t$ increases, the interface moves 
so that the fraction of the phase with lower free energy increases. 
The melting pressure $P_m$ can be determined as the point 
where the sign of $d\phi/dt$ in the steady state changes, 
since positive and negative values of $d\phi/dt$ indicate 
that crystallization and melting occur at the interface, respectively. 
This method requires a relatively short-time simulation 
compared to the equilibrium method 
and enables us to treat larger systems 
and reduce the finite-size effect. 
Figure \ref{fig:dphidt-P} shows the pressure dependence 
of $d\phi/dt$ and obtained $P_m(\Delta)$ 
gives the phase boundary between fluid and crystal phases 
in the inset of Fig.~\ref{fig:PD}.

By this steady interface motion, 
we can compare {\it equilibrium} stability of FCC and fluid states. 
There remains a possibility, however, 
that there can be more stable state, 
such as other types of crystal structure. 
But we expect that it is not the case for polydispersity, $\Delta<0.088$.

\begin{figure}
\begin{center}
\includegraphics[trim=20 250 0 -240,scale=0.30,clip]{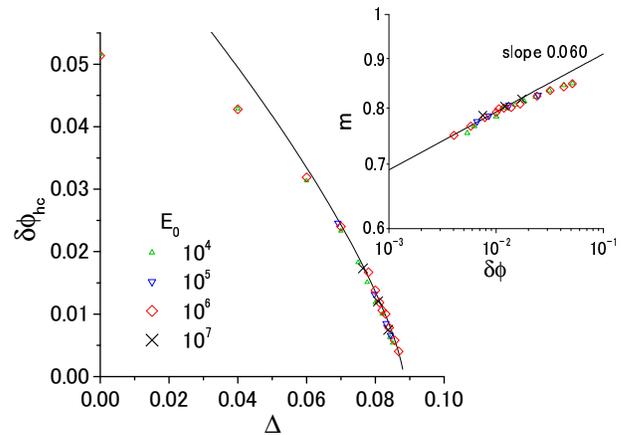}
\end{center}
\vspace{-5mm}
\caption{\label{fig:Deltaphi-Delta}
(color online) 
Polydispersity dependence of the density gap 
between the fluid and crystal states on the melting line. 
The results for different Young's moduli are shown together 
but very little difference is observed. 
The solid curve denotes 
$\delta \phi = 0.45 \times (0.088 - \Delta)^{0.73}$. 
(inset) Log-log plot of the relation 
between the density gap and the crystalline order parameter 
on the melting line.
}
\end{figure}

Performing additional equilibrium simulations at these $P_m(\Delta)$, 
we eventually obtain $\Delta$-dependence 
of the $\delta \phi$ shown in Fig.~\ref{fig:Deltaphi-Delta}. 
This order parameter approaches zero as 
$\delta\phi \propto (\Delta_c - \Delta)^\beta$ 
with $\Delta_c = 0.088(2)$ and $\beta = 0.7(2)$. 
The critical pressure $P_c = 3.0(2)$ 
and the critical packing fraction $\phi_c = 0.576(4)$ 
are also obtained. 
We also calculate the FCC order parameter of the crystal state;  
$m = \overline{
\cos[\mbf{K}_\mrm{FCC} \cdot (\qq_i(t)-\qq_i(0)) ] 
}$, where $\mbf{K}_\mrm{FCC}$ is the fundamental reciprocal vector 
of the FCC crystal. 
The inset of Fig.~\ref{fig:Deltaphi-Delta} indicates a power-law; 
$m \propto \delta \phi(\Delta)^{\beta_m/\beta} 
\propto (\Delta_c - \Delta)^{\beta_m}$ 
with $\beta_m = 0.04(1)$. 
The range of observed value of $m$ is, however, too narrow 
to conclude the existence of the power-law.


\section{Glass transition}

We next investigate the transition between the fluid and glass phases 
by scanning $P$ at fixed $\Delta (> \Delta_c)$. 
Significant change is observed around a certain threshold $P_g(\Delta)$; 
the mobility of particles markedly decreases approaching $P_g$, 
which denotes the glass transition point. 

\begin{figure}
\begin{center}
\includegraphics[trim=20 230 100 -250,scale=0.36,clip]{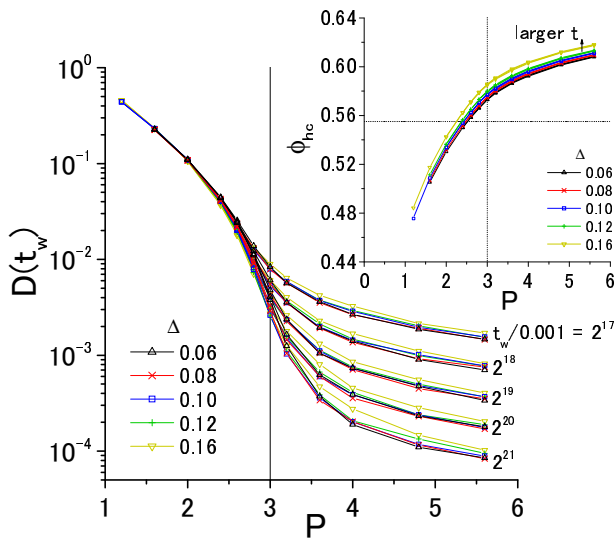}
\end{center}
\vspace{-5mm}
\caption{\label{fig:Diff-P}
(color online) 
Pressure dependence of the diffusion constant 
for fixed polydispersity. 
The data for various waiting times 
are plotted together to show the aging behavior. 
The initial state is random packing with packing fraction 0.50. 
(inset) Pressure dependence of the packing fraction 
for fixed polydispersity. 
For each $\Delta$, we show the data at three times, 
$t/0.001 = 2^{20}, 2^{21},$ and $2^{22}$ 
to show good convergence. 
}
\end{figure}

Figure~\ref{fig:Diff-P} shows the $P$-dependence of the diffusion constant, 
$D(t_w) = \overline{ |\qq_i(2t_w) - \qq_i(t_w)|^2} / t_w$, 
with waiting time $t_w$ under the random initial condition. 
Here we make the time interval to measure the displacement 
equivalent with $t_w$ so that only single time scale is introduced. 
While the $D(t_w)$ converges to a certain equilibrium value 
by increasing $t_w$ for $P<P_g \approx 3.0$, 
the relaxation is so slow for $P>P_g$ that equilibrium state cannot 
be obtained for used values of $t_w$. 
Instead, we remarks on the aging property, 
i.e., the persistent waiting-time dependence; 
$D(t_w)$ continues to decrease with $t_w$, 
roughly in a power law, above $P_g$. 
This indicates that as the relaxation proceeds, 
the system becomes trapped 
in an increasingly stable metastable state 
and the dynamics becomes slower.

As clearly observed in Fig.~\ref{fig:Diff-P}, 
the $D(t_w)$ vs $P$ curves hardly depend on 
$\Delta$ for $\Delta \le 0.12$. 
Thus $P_g(\Delta)$ is also independent of $\Delta$ 
and equals to 3.0, similar to the value of $P_c$. 
In general, the effect of polydispersity is small 
except in the crystal phase. 
In addition, almost the same behavior is observed 
even in the subcritical region ($\Delta < \Delta_c$), 
as a supersaturation phenomena \cite{Zaccarelli09}. 
Any sign of crystal nucleation is not observed for $\Delta \ge 0.60$. 
It is known that polydispersity drastically 
reduces the nucleation rate of the crystal \cite{Auer01}. 
In the inset of Fig.~\ref{fig:Diff-P}, 
$\phihc$ is plotted with respect to $P$. 
The packing fraction also has little dependence on $\Delta$ 
(slightly increases with $\Delta$) both in the fluid and glass phases. 
Therefore, the glass transition density 
$\phi_g(\Delta) \equiv \phi \left(P_g(\Delta) \right)$ 
also has little dependence on $\Delta$ 
and $\phi_g \approx 0.57 \approx \phi_c$. 
The extrapolated value, $\phi_g(\Delta \rightarrow 0)\approx0.57$, 
agrees with the value, $\phi_d \approx 0.58$, 
predicted by mode-coupling theory \cite{vanMegen93} or mean field theory, 
which corresponds to the appearance of the exponentially many 
metastable states in the fluid \cite{Parisi08}.

Above the threshold pressure $P_g$, 
$\phihc$ gradually approaches the random close packing (RCP) fraction 
$\phi_\mrm{RCP}(\Delta)$ \cite{Scott69}, 
which equals 0.635 for the monodisperse $(\Delta=0)$ system 
and increases very slowly with $\Delta$ \cite{Rintoul96}. 

\section{Crystal-amorphous transition}

\begin{figure}
\begin{center}
\includegraphics[trim=0 250 60 -225,scale=0.28,clip]{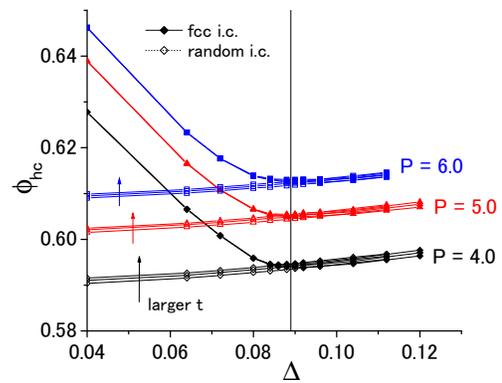}
\end{center}
\vspace{-5mm}
\caption{\label{fig:phi-Delta}
(color online) 
Polydispersity dependence of the packing fraction at fixed pressure. 
The data for FCC and random initial configurations are plotted together, 
which coincide for large $\Delta$. 
In each case, we show the data at three times, 
$t/0.001 = 2^{16}, 2^{17}$, and $2^{18}$. 
}
\end{figure}

Finally, we consider the transition 
between the crystal and glass states, 
which is driven by sweeping $\Delta$ at fixed $P(>P_c)$. 
In Fig.~\ref{fig:phi-Delta}, we plot the $\Delta$-dependence 
of $\phihc$, estimated by simulations 
under both crystal and random (glass) initial conditions. 
Both the crystal and glass states are (meta)stable for a long time 
in the high-pressure region 
since there is too little free volume for the local structures 
to reconfigurate. 
The packing fraction of the crystal is larger 
than that of the glass for small $\Delta$ 
but they becomes indistinguishable 
above a certain threshold $\Delta_\mrm{am}(P)$. 
This threshold appears to be universal, 
i.e., it hardly depends on $P$. 
In addition, its value is very similar to $\Delta_c \approx 0.088$. 
The density difference continuously decreases to zero 
as $\Delta$ approaches $\Delta_\mrm{am}$. 
Similar behavior is observed for the crystalline order parameter 
under the FCC initial condition.

\section{Conclusions}

In summary, 
we investigated the dynamical transitions 
of a polydisperse elastic sphere system by MD simulations 
by remarking on nonequilibrium states including metastable states. 
The obtained transition lines with respect to 
the packing fraction and polydispersity are summarized 
in Fig.~\ref{fig:PD}. 
It was confirmed that the first-order transition 
between the fluid and crystal phases 
terminates at the critical point $(\phi_c, \Delta_c)$ 
and that the other two phase boundaries begin from the critical point 
to surround the glass phase. 
The glass state has intermediate properties 
between those of the fluids and crystal states; 
it exhibits temporal freezing but does not have periodic order. 
The fluid-glass and crystal-glass boundaries can be 
drawn in a surprisingly simple way 
and are expressed as $P \simeq P_c$ (or $\phi \simeq \phi_c$) 
and $\Delta \simeq \Delta_c$, respectively. 
The glass transition line passes 
through the critical point, which is reasonable 
because the continuous breakdown of the crystal requires 
marginal fluidity at the critical point.  
While softening of the interaction potential 
will not make essential change in the phase behavior, 
that of a system with attractive interactions 
is an interesting open problem.

The transition between the fluid and glass states is 
not considered to be an equilibrium transition 
but a dynamical one 
since the static quantities do not exhibit any singular behavior 
and the transition line is elongated into the crystal phase in equilibrium.  
This continuous relationship from the supercritical region 
to the super-saturating monodisperse system 
suggests the equivalence of the dynamical glass transitions 
in the monodisperse and polydisperse systems. 
Let us note that the critical packing fraction is close 
to the random loose parking (RLP) fraction, 
$\phi_\mathrm{RLP} \approx 0.56$, 
which is considered to be the minimum packing fraction 
required to maintain the internal stress for highly frictional particles 
\cite{Onoda90}. 
This coincidence seems natural considering that RLP gives a criterion 
related to the excluded volume effect.  
The free volume of particles is very small above $\phi_\mrm{RLP}$ 
and diffusion is highly suppressed.

Let us consider the meaning of the boundary 
between the crystal and glass states. 
The transitions at this boundary are continuous in terms of density, 
in contrast to those of fluid-crystal boundary in the subcritical region. 
It is natural that the first-order transition line 
and a continuous transition line should meet at the multicritical point. 
But this is conflict with the prediction of first order transition 
by Bartlett and Warren \cite{Bartlett99}. 
Our nonequilibrium analysis cannot eliminate the possibility that 
the crystal phase is metastable below $\Delta_\mrm{am}$, 
which is estimated by nonequilibrium simulations, 
and first-order transition occurs at $\Delta < \Delta_\mrm{am}$. 
We wonder, however,  whether a mean-field-like approach in the DFT scheme 
can treat the state around the terminal point, 
where the fluid exhibits singular behavior in dynamics 
and the periodicity of the crystal is damaged. 
Our numerical result suggests the possibility that criticality remains.




\acknowledgments
This work is supported by KAUST GRP(KUK-I1-005-04) 
and Grants-in-Aid for Scientific Research (Contracts No.~19740235).



\end{document}